# Reengineering observatory operations for the time domain

Robert L. Seaman[a][a], W. Thomas Vestrand[b], Frederic V. Hessman[c]

[a]National Optical Astronomy Observatory, 950 N. Cherry Ave., Tucson, AZ, USA 85719;
[b]Los Alamos National Laboratory, USA; [c]Georg-August-Universität, Göttingen, Germany

## ABSTRACT

Observatories are complex scientific and technical institutions serving diverse users and purposes. Their telescopes, instruments, software, and human resources engage in interwoven workflows over a broad range of timescales. These workflows have been tuned to be responsive to concepts of observatory operations that were applicable when various assets were commissioned, years or decades in the past. The astronomical community is entering an era of rapid change increasingly characterized by large time domain surveys, robotic telescopes and automated infrastructures, and – most significantly – of operating modes and scientific consortia that span our individual facilities, joining them into complex network entities.

Observatories must adapt and numerous initiatives are in progress that focus on redesigning individual components out of the astronomical toolkit. New instrumentation is both more capable and more complex than ever, and even simple instruments may have powerful observation scripting capabilities. Remote and queue observing modes are now widespread. Data archives are becoming ubiquitous. Virtual observatory standards and protocols and astroinformatics data-mining techniques layered on these are areas of active development. Indeed, new large-aperture ground-based telescopes may be as expensive as space missions and have similarly formal project management processes and large data management requirements.

This piecewise approach is not enough. Whatever challenges of funding or politics facing the national and international astronomical communities it will be more efficient – scientifically as well as in the usual figures of merit of cost, schedule, performance, and risks – to explicitly address the systems engineering of the astronomical community as a whole.

**Keywords:** observatory operations, time domain astronomy, transient events, VOEvent, robotic telescopes

## 1. INTERCONNECTIONS

Henry David Thoreau said, "*The question is not what you look at, but what you see.*"[b] On the other hand, it is safe to say that the consensus in the scientific community would be that both what you look at, and how you go about doing so, greatly influence what you see. In particular, more and more astronomical research programs involve coordinating observations at multiple facilities. This is especially true in the time domain, for instance when following up discoveries of celestial transient events from large ground-based surveys or from space-based gamma-ray telescopes. It would be far better to implement capabilities for coordinated observing modes as a coherent system design spanning the community then to rely on ad hoc last-minute arrangements.

This conference is named "*Observatory Operations: Strategies, Processes, and Systems V*". It is interesting to note that every word in that title is plural – except for the one that defines it most, "Observatory". To operate a single observatory is indeed a challenge, and the concepts of operations for our observatories span not just remote mountaintops and orbiting platforms, but also balloons and airplanes, experiments under the waves and deep beneath the Earth, and on and under the Antarctic ice cap. These telescopes and instruments operate using remarkably diverse notions of the word "observation" and their detectors are sensitive to all parts of the electromagnetic spectrum, to esoteric particles and gravitational waves. But they are all designed to answer questions posed by the universe – and those questions are often contingent on the answers derived from other surveys and experiments conducted at other observatories.

---

[a] seaman@noao.edu

[b] Thoreau, H. D., "Journal, 5 August 1851", https://www.walden.org/Library/Quotations/Observation

That the time has come for the astronomical community to think about its observatories in the plural is not a revolutionary concept. In essence this is the argument behind the "U.S. Ground-Based Optical/Infrared (O/IR) System" as motivated by the 2000 U.S. Decadal Survey.[c1] Indeed, observational astronomy is conducted via a ramified system-of-systems of national and international, public and private, multi-wavelength, multi-messenger observatories providing access to telescopes of all apertures and to instrumentation of diverse capabilities. A program of research begun at one facility often spawns follow-up investigations at numerous others carried out by the original team or by new researchers and responsive over both very short latency timescales or extending for years or decades into the future.

## 1.1 Motivating community infrastructure

The astronomical community has a long history of shared effort on common standards and infrastructure. FITS[d2] has been our common data interchange format for more than three decades. Telegrams conveying news of celestial events were among the first sent by the Transatlantic Cable in the 19th century.[3] Multiple telescopes often share power and networking, mirror-aluminization, dormitory, and lunchroom facilities, *etc.*, on a single mountaintop. Many such examples exist. But there are many more examples of duplicated efforts and manual workarounds, rather than of coordinating efforts and facilities for mutual benefit. This is especially true the more tightly coupled the science use cases become. It is easiest to see this coupling in the time domain, where hours or days lost to some ad hoc email exchange of source details may squander an observing opportunity, but is true whenever human effort must be expended to check and recheck target lists to avoid typos, or when two teams duplicate efforts rather than spread them for maximum observing efficiency across a range of targets.

In the 1990s observatories pursued improved efficiency through remote observing[4] and "new observing modes"[5] such as service observing and observation queues. New operational paradigms have been coupled with robotic telescopes and autonomous instrumentation to optimize observing efficiencies for specific science programs at particular observatories. These efforts have rarely extended beyond more than one observatory. Recent years have seen the commissioning of networks of heterogeneous telescopes[6] or more self-consistent networks[e], but these fall short of community-wide solutions.

The broader context is the emergence of very diverse models of telescope allocation. At the public facilities there are a variety of TAC (Telescope Allocation Committee) processes. Major private facilities generally reserve a significant fraction of the time for guaranteed staff use. Highly integrated projects like the Palomar Transient Factory (PTF), and its successor, the Zwicky Transient Factory (ZTF)[f] may allocate an entire telescope to one activity. Then there are the various expressions of the role of Director's time, such as produced the Hubble Ultra Deep Field[7]. These internal policies have to be balanced in some fashion against external commitments. Observatories have partners who are granted large amounts of time (e.g., the Dark Energy Survey at NOAO), medium amounts for smaller surveys or targeted synoptic programs, or at the other end of the spectrum, fractional nights for Target-of-Opportunity (ToO) follow-up. During any given observing semester an observatory may be cooperating with several others through carefully negotiated memoranda-of-understanding (MOUs). Recent years have seen increasing collaboration of astronomers with researchers in other fields such as particle physics and computer science. In the absence of coherently designed system interfaces and deployed community-wide infrastructure, each of these many variations must be negotiated and implemented separately, not only losing out on opportunities for efficiencies of reuse and scalability, but often conflicting with each other and with more traditional classical observing modes.

After a TAC (or equivalent process) allocates telescope time to different science programs, it is a familiar notion that a single major telescope may support many different instruments that are timeshared via a scheduling process that implements the TAC's decisions. Different telescopes provide differing levels of efficiency in swapping instruments. Indeed, often instruments are shared between multiple telescopes requiring significant re-commissioning activities. Different observatories support more flexibility or significantly less flexibility in updating observing schedules after the fact. When an instrument is tied to a telescope or to an observing program, the telescope is productive only if there's a constant stream of productive observers or a constant stream of service observations from the pool of astronomers with access, or alternately to rapid changes in instrumentation, which adversely affects efficiency.

---

[c] "Astronomy and Astrophysics in the New Millennium", http://www.nap.edu/catalog.php?record_id=9839
[d] http://fits.gsfc.nasa.gov/fits_standard.html
[e] http://lcogt.net
[f] http://www.ptf.caltech.edu

On the other hand, there is less need to change instrumentation if the larger network contains diverse instrumentation that can be shared. Similarly complications and efficiencies are traded off at all levels of observatory operations when viewed as a community-wide system of facilities. As the next section will show, there are many ways in which astronomical facilities are increasingly interoperating. While some commonalities of purpose naturally evolve, others would benefit from coherent system engineering and community-wide infrastructure development.

## 2. EXAMPLES OF COMMUNITY SYSTEM ISSUES FROM THIS CONFERENCE

The contributions to the agenda of any conference provide a snapshot of issues that are important to the astronomical community at any given moment in time. This is especially true of SPIE Observatory Operations, one of the few venues in which those responsible for operating the great diversity of astronomical facilities can get together to discuss the many common or unique issues that define observational astronomy as a scientific and technical pursuit. The remainder of this section is organized around quotes from about a third of the conference contributions, chosen to illuminate a variety of challenges and opportunities from addressing the plurality of observatories as the complex system that it is in fact. Only a loose order is imposed and the reader will likely think of other issues that were simply not addressed at this year's meeting. Quotes are in *blue italics*; **key points are bold**; commentary on each paper is contained in a single paragraph.

From Stoehr, *et. al.*[8] [9149-1] *"We argue that in the future observatories will compete for astronomers to work with their data, that observatories will have to reorient themselves from providing good data only to providing an excellent end-to-end user-experience with all its implications, that science-grade data-reduction pipelines will become an integral part of the design of a new observatory or instrument and that* **all this evolution will have a deep impact on how astronomers will do science.**" This is as cogent a summary of current trends as one will find. An implication of providing an excellent end-to-end user-experience is ensuring that users of multiple facilities can reuse their hard learned experience between observatories, and can easily combine all phases of observational astronomy from the proposal process, to phased observing run preparation, to scheduling and carrying out observations, to the joint interpretation of the raw and pipeline-reduced data products that originate from diverse instrumentation and archives. *"The success of ALMA and of any other astronomical facility is measured by the scientific output of the community. As thus by construction* **the facility cannot assure their own success directly**, *the way to improve success is to render the end-to-end user-experience for PIs and archival researchers as perfect as possible."* This is not only the way to enhance 'customer satisfaction', it also addresses the fundamental bottom line of the individual facilities. *"The way astronomy is done has changed dramatically over time. ... ALMA will produce about the same amount of data in one year as ESO's telescopes have produced in its first 50 years. But ESO, too, will soon produce data at the same rate! ... Whereas data will scale exponentially astronomers will not. Therefore the bytes per astronomer do scale exponentially. Our prediction is that whereas now astronomers are competing for observing time, in the future, observatories will be competing for astronomers."* The argument is not just that these are good ideas, rather that these are necessary changes that will be required by the evolving scales and complexities of instrumentation and data.

Crabtree[9] [9149-10] says *"The primary scientific output from an astronomical telescope is the collection of papers published in refereed journals. A telescope's productivity is measured by the number of papers published which are based upon data taken with the telescope. ... When, as often is the case, a paper is counted by more than one observatory I give each observatory full credit for the paper. Division of the credit (citations) between different telescopes is subjective... Some of the observatories in this study consist of multiple telescopes,* e.g., *Keck. In these cases I simply divided the number of papers by the number of telescopes to calculate the productivity."* It is often the case, and will become increasingly so, that individual papers rely on observations taken at multiple observatories. This is already complicating the compilation of statistics used to measure the productivity of different facilities. As data sets become more intertwined – from being combined after the fact, to being contingent follow-up observations or even simultaneous co-observations – it will require not just more careful accounting, but entirely new ways of doing so beyond tallying papers in multiple columns on the one hand versus splitting the difference on the other.

From Saha, *et. al.*[10] [9149-7] *"The Arizona-NOAO Temporal Analysis and Response to Events System (ANTARES) is a joint project of the National Optical Astronomy Observatory and the Department of Computer Science at the University of Arizona. The goal is to build the software infrastructure necessary to process and filter alerts produced by time-domain surveys, with the ultimate source of such alerts being the Large Synoptic Survey Telescope (LSST). The ANTARES broker will add value to alerts by annotating them with information from external sources such as previous surveys from across the electromagnetic spectrum. In addition, the temporal history of annotated alerts will provide further annotation for analysis."* The scientific interpretation of astronomical data, in particular of reports of celestial transient events, is highly dependent on the context provided by survey catalogs and multi-bandpass sky maps, *etc*. When data are static or alerts are rare, this context can be explored manually by the astronomer. In the future this human-mediated data interpretation will be less-and-less practical and **automated power tools will be required to make sense of the data rapidly enough to make useful follow-up observations practical**. A community resource such as ANTARES is particularly valuable since its outputs can be fed to multiple subscribing observatories and astronomers, benefiting from a network multiplying effect.

Fitzpatrick, *et. al.*[11] [9149-65] states *"[T]he role of the astronomer is changing from a traditional approach of single investigators or small groups collecting, calibrating, and analyzing small sets of data, to one of working in big teams to run surveys that generate large collections of data used by broad swaths of the community."* What is true for event data streams is increasingly true for all astronomical data sets. **The efficient handling of big, diverse data sets shared among large multi-institutional teams requires a change of data-handling paradigm**.

Diehl, *et. al.*[12] [9149-31] *"The Dark Energy Survey (DES) is an international collaboration, with over 200 scientists from over 20 institutions in the US, the UK, Spain, Brazil, Switzerland, and Germany. …. The observers maintain commentary and notes in an electronic logbook. … A series of automatically generated plots and statistics follows…"* Large collaborations require power tools beginning at the telescope(s), not only downstream in data management systems. *"The handshake between the telescope controls and the camera controls can be improved to shave a couple of seconds off the time between shutter-close and shutter-open when there has been no slew between exposures."* The economies of scale with large projects place a large premium on even overtly small improvements in efficiency. Two seconds per exposure times hundreds of exposures per night represents dozens of additional exposures that can be taken every night. This is especially true when the scientific output of multiple observatories is combined since the dimensionality of the parameter space may leverage the joint efficiencies as the squared or cubed power.

From Boroson, *et. al.*[13] [9149-50] *"The network has been designed and built to allow regular monitoring of time-variable or moving objects with any cadence, as well as rapid response to external alerts. ... The unique attributes of the LCOGT network make it different enough from any existing facility that alternative approaches to optimize science productivity can be considered."* And Pickles, *et. al.*[14] [9149-38] says *"We describe the operational capabilities of the Las Cumbres Observatory Global Telescope Network. We summarize our hardware and software for maintaining and monitoring network health. We focus on methodologies to utilize the automated system to monitor availability of sites, instruments and telescopes, to monitor performance, permit automatic recovery, and provide automatic error reporting."* LCOGT is indeed a unique and exciting new facility. The slogan of SUN Microsystems was 'The Network is the Computer' and a slogan for LCOGT could well be that **The Network is the Telescope**. Its model of requesting observations from a telescope network rather than a particular telescope is proof-of-concept for sharing facilities community-wide. On the other hand such flexibility exacts a cost in requiring adherence to a common set of standards, to deploying infrastructure throughout the network, and to subordinating local control to a distributed model.

Saunders, *et. al.*[15] [9149-14] states *"LCOGT is developing a worldwide network of fully robotic optical telescopes dedicated to time-domain astronomy. Observatory automation, longitudinal spacing of the sites, and a centralized network scheduler enable **a range of observing modes impossible with traditional manual observing from a single location**."* Coordinating observing facilities does not just offer scheduling efficiencies, it enables new kinds of observing. This is true for a network of similar aperture telescopes with standardized instrumentation, and can be even more true for heterogeneous networks containing diverse apertures and instrumentation. *"Re-scheduling never pre-empts a block that is already executing at a telescope at scheduling time; interrupting such blocks is inefficient and caused significant 'thrashing'. The only exception to this rule are requests flagged as time critical, called Target-of-Opportunity (ToO). These requests require immediate placement at the first available telescope. As long as ToO requests comprise only a small fraction of the total request pool, such interruptions are tolerable."* But coordinating facilities comes with a cost expressed in the 'rules of engagement'. Each networked telescope must obey common scheduling policies of one sort or another. If there are exceptions, they must remain infrequent and carefully managed. *"A planned schedule has many advantages over a simple dynamic queue, but in a planned schedule such overhead estimates need to be as accurate as possible. Miscalculations in the overheads lead either to schedule inefficiency (if the estimates are too large), or impossible schedules (if the estimates are too small). Because some components of the overhead are either difficult to model (*e.g.*, slew time, which depends on the choice of preceding observation) or semi-stochastic (*e.g. delays introduced by axis unwrapping, or CPU load at site), we have opted for somewhat conservative overhead estimates. This will certainly have an impact on our overall open-shutter efficiency. Quantifying and refining the overheads remains an ongoing challenge."* **Coordinated observing facilities will require that accurate estimates of their multi-

**dimensional capabilities be made public**. An instrument must own up to its overhead. A telescope cannot hide its logistical complications. *"Users need accurate feedback. ... All aspects of scheduling need clear metrics. ... The definition of 'successful' is nuanced, and needs work."* And the concept of success must be expressed in terms understood by all. In general this will be tied to scientific success, and one can expect counter-intuitive results from combining the capabilities of major and minor facilities in diverse ways.

From Gonté and Smette[16] [9149-39] *"The Paranal Observatory instrumentation uses virtually all the possible technologies developed by the astronomy community during the last 20 years. Spanning a wavelength coverage from 315 nm to more than 20 microns, it includes adaptive optics and interferometric system for high spatial resolution imaging and spectroscopy, large field of view imager, Multi Object Spectrometer, Integral Field Unit, High resolution spectrometers etc… Each instrument can be considered as a separate entity but **the instrumentation can be viewed also as a global system** implemented over the years on Paranal."* A telescope is a system of multiple instruments. A mountaintop is a system of multiple telescopes. An observatory may manage several mountaintops / spacecraft. Individual observatories join into partnerships or classes of different sorts: public, private, optical, IR, radio, ground-based, space-based, national, international, *etc*. And these classes of facilities join into the world-wide astronomical community. With each increase of scale some level of coordination and focus is lost. Reintroducing standards and common infrastructure that span the many levels from community to instrument shutter – even on a modest pragmatic scale – can provide improvements of observing efficiency and access to science modes that can be achieved no other way.

Chandler and Butler[17] [9149-43] says *"There were two main impacts on astronomical observing of having concurrent construction, commissioning, and operations activities: (1) a deduction in time available for astronomy, and (2) increased risk of failure because of new software and hardware being deployed during construction activity during the work day."* The world-wide astronomical community – really now a Solar-system-wide astronomical system extending outward to the Voyager 1 spacecraft at the edge of interstellar space – is constantly under construction. This will also be true of individual missions or ground-based sites during some periods of time. The same risks apply to coordinated observations, archival data searches, shared software assets, optics laboratories and other common infrastructure that are similarly always evolving. *"**Time-critical observations**, which used to be labor intensive and disruptive to schedule, **became much more easily accommodated by the move to dynamic scheduling**. Indeed, if an observation needs to be observed within a few hours of a trigger, experienced observers are allowed to self-approve their SBs…"* Time-critical observations, or any observations that cross outside the normal time allocation procedures for a particular facility, imply that an evolution may be needed in observatory policies and procedures.

Walther, Dempsey and Campbell[18] [9149-93] *"[I]t was decided to take advantage of the hours between when the telescope operator leaves the telescope and when the day crew arrives. ... This paper describes the hardware changes necessary to implement remote observing at JCMT. It also describes the software needed for remote, fail safe, operation of the telescope."* And on the other hand, an explicit policy change – here, to extend observing hours – has implications for both hardware and software infrastructure, for operational procedures, safety issues, training, *etc*. *"Since these Extended Operators are not expert telescope operators, the system was simplified as much as possible, but some training was necessary and proper checklists are essential. ... The EO is not really expected to be capable of solving a lot of faults."* Such after-the-fact changes are likely to be limited in scope since system requirements, stated or unstated, will be violated in the refactoring of the system. Multiple tiers of service may be created. *"[T]he safety of uninvited visitors to the telescope must be considered also. ... [T]he EO…must click a button on the Hilo Handshake Status GUI…every 30 minutes."* Previously solved problems have to be revisited. Decisions such as the safe operating parameters of a staffed observatory no longer apply when the staff is no longer present. *"As soon as the University heard that JCMT was going to spend more time observing, they wanted some of that time."* And previously negotiated agreements regarding resource allocation or the division of institutional responsibilities, etc., may need to be reopened.

From Swindell, *et. al.*[19] [9149-75] *"We describe a complex process needed to turn an existing, old, operational observatory – The Steward Observatory's 61" Kuiper Telescope – into a fully autonomous system… Based on environmental sensors and internal calculations, observatory control software assumes responsibility for opening and closing the observatory, pointing the telescope and commanding camera(s) and other observatory hardware. **This truly enables new science**, ranging from fast target of opportunity observations to long duration surveys, done without any need for expensive human oversight."* Ideally autonomous facilities would be built with these capabilities planned from scratch. This has been rarely true to date. Even a single-telescope observatory is a complex entity, and many separate subsystems need to be updated before the total system can be called fully autonomous. Such a project needs to be justified on the basis of specific new science use cases that can be addressed using the new capabilities. *"The 61" is liked by present-day observers because of its fine optics as well as its cozy dormitory reminiscent of a ski lodge and the spectacular views it offers."* Don't discount the sociology of astronomy. Observing on a remote mountaintop is one of the highlights of being an astronomer. Even if the entire infrastructure of astronomy could be automated, some human-steered telescopes would likely remain to provide training for the next generation, or as a platform for developing new concepts for instrumentation, or to carry out whatever irreducible minimum of classical observing programs is deemed appropriate. *"RTS2 drivers will be written for every device present at the observatory. For basic operations, telescope and camera drivers are enough – those will allow RTS2 to handle scheduling and demonstrate its functionality. For complete autonomous operations, at least the dome, safety systems (weather and power sensors), filter and focuser (if present) drivers should be running."* Autonomous features can be addressed as different levels of service. An equatorial telescope with a clock drive could be described as "autonomous". Add a star tracker and it will follow a target all night long. Most CCD data acquisition software provide a feature for taking multi-exposure observing sequences. Data flow automatically into archives. Each of these features is a rung on the autonomous ladder. *"RTS2 driver functions are encapsulated inside a class. ... Other methods may be used to react on data and events gathered by the daemon. ... Each driver is a standalone program. ... The RTS2 library is a single thread. ... [I]t is able to react to external events, which happens during some other action performed by the driver. It is left to the driver author to write drivers which will provide adequate response times."* Software details are often invisible to the astronomer, but ultimately govern the capabilities and logistics of the system. Even a simple data acquisition system has vast numbers of degrees of freedom and observers will often make assumptions as to the behavior of the system by analogy with instrumentation layered on some completely different software concept, *e.g.*, an object-oriented application might be compared to the behavior of a traditional procedural program. This is often visible in timing issues throughout a exposure cycle when real-time microcode hands off to event-driven routines waiting on the settle time of telescope servo loops. *"[Central] logging is critical for development, testing and debugging of the fully autonomous system. Without night logs, it will be impossible to understand why the system behaves the way it does. Primarily, the failed device is responsible for recovery from the failure on its own. If the recovery from failure requires any other device to perform certain commands, the best option is to use an external script to recover from failure."* The normal state of affairs on a mountaintop is the handling of exceptions and edge cases. When observatories interoperate, the edge cases are compounded. Appropriate logging is critical, especially if power users from other observatories are involved. Autonomous operations implies autonomous recovery from errors, autonomous stowing of the telescope when the weather changes, detection or inference of astronomical twilight, autonomous calibrations, etc. *"[I]t is difficult to keep a single observation scheduling productive. ... We would like to adopt an approach similar to the 1.2m telescope at the nearby FLWO – meta queue scheduling coupled with merit function backup."* Robotic observatories have come up with numerous clever schemes for queue scheduling handling ToO interrupts. Care will be needed to support multiple schemes while permitting flexibility in joint observation modes.

Bauman, *et. al.*[20] [9149-55] says *"CFHT's decision to move away from classical observing prompted the development of a remote observing environment… A comprehensive feasibility study was conducted to determine the options available to achieve remote operations of the observatory dome drive system."* **Requirements for new capabilities will flow through all aspects of the operation of the observatory**, even basic functionality such as operation of the dome, delivery of cryogens, non-interruptible power, remote console access to reboot computers, an automated response to adverse changes in the weather, *etc.* Each of these many areas may influence others and all will benefit from comprehensive planning, ideally in the larger context of an entire observatory or community, if only to avoid re-inventing the wheel.

Dodd[21] [9149-40] states *"This paper discusses similarities and differences in operations and management between two NASA astrophysics missions…The mission size, cost trajectory, and management structure are different between Spitzer and NuSTAR, yet there are similarities in operations and management approaches."* Diverse engineering problems may result in similar solutions – or overtly similar problems may diverge into quite different solution spaces.

From Forster, *et. al.*[22] [9149-27] *"NuSTAR complements observations using current soft X-ray observatories like Chandra, XMM-Newton, Suaku, and Swift and is a bridge to low resolution non-focusing of coded mask instruments operated in the hard X-ray band by INTEGRAL, Fermi, MAXI, and Swift-BAT. The capabilities of NuSTAR are being used to break model degeneracy within interpretations from soft X-ray observations alone. This capability was understood before launch. ...* **The level of coordination with other observatories**, *including ground-based observatories like HESS, MAGIC, and VERITAS,* **is unprecedented for a new mission; with more then 15% of all NuSTAR observations requiring some coordination of scheduling**.*"* Coordinated observing modes are increasingly being designed as part of the base mission of different facilities. This is especially true of space-based telescopes, but community follow-up is also fundamental to various science use cases for new ground-based telescopes such as LSST. *"Scheduling constraints for NuSTAR are mild in comparison with other space observatories and so the NuSTAR schedule will most often follow the timing of observations for other observatories."* An asymmetry of telescope capabilities will often be evident when coordinating observations. This needs to be accommodated in any system-of-systems. *"NuSTAR is designed to be able to access 80% of the sky at any given time and this makes it a powerful telescope for Target of Opportunity (ToO) investigations. [R]esponse times to ToOs [are] limited by staffing…(standard business hours) and the availability of ground station or TDRSS contacts."* Strong inherent capabilities will often be limited by purely logistical constraints.

Altmann, *et. al.*[23] [9149-25] says *"To fully achieve the ambitious goals of the [Gaia] mission, and to completely eliminate effects such as aberration, we must know the position and velocity vectors of the spacecraft as it orbits the Lagrange point to an accuracy greater than can be obtained by traditional radar techniques, leading to the decision to conduct astrometric observations of the Gaia satellite itself from the ground. Therefore the Ground Based Optical Tracking (GBOT) project was formed and a small worldwide network using 1-2 m telescopes established in order to obtain one measurement per day of a precision/accuracy of 20 mas."* All observatories, whether space or ground-based, provide viewpoints unique in time and space. For some science programs the topocentric or orbital coordinates need only be approximately known. For others, for example asteroid occultations, very precise coordinates relative to a formal geographic datum must be known. Astrometric accuracy is hard to add back in after the fact. Engineering requirements on such capabilities trace directly to the kind of science that will later be possible with a particular facility. This is an excellent example where **cooperation can result in substantial but indirect benefits to the ultimate science**.

Gaug, *et. al.*[24] [9149-45] says *"The absolute calibration of Cherenkov telescopes can be cross-checked with space instruments such as Fermi-LAT. … We have presented our plans to ensure that the telescopes, as well as the atmosphere, both of which are an essential part of the detection principle, are always understood to required precision."* Though not usually presented as such, any mention within the astronomy literature of calibrating data from one telescope versus that from another, is really stating a requirement on a system larger than both. Similarly, the Earth's atmosphere is not just an impediment to observations, it is an intrinsic part of all ground-based observing systems.

From Jankowsky and Wagner[25] [9149-70] *"Imaging Atmospheric Cherenkov Telescopes (IACT) utilize the Earth atmosphere as calorimeter of the detector system. … The energy of the primary particle is then reconstructed via comparison with Monte Carlo simulations, which is complicated by changes in atmospheric conditions, as they directly influence the energy threshold of the measurements."* This recognition that the atmosphere is part of the detection process is integral to the concept of a Cherenkov Telescope (and other instruments such as speckle interferometry), but remains true of any observation made from the ground. In other words, the roiling atmosphere is part of the world-wide astronomical observing system. And Jankowsky and Wagner[26] [9149-71] state *"The High Energy Stereoscopic System (H.E.S.S.) is an array of five Imaging Atmospheric Cherenkov Telescopes (IACT) located in the Khomas Highlands of Namibia."* It bears stating that few other scientific enterprises enjoy the diversity of remote siting as does astronomy. This is a fact fundamental to this extensive system-of-systems, not a complication to be handled separately. *"It is sensitive to Very High Energy (VHE) gamma rays between hundreds of GeV to tens of TeV. … Forming part of H.E.S.S., the Automatic Telescope for Optical Monitoring (ATOM) is an optical 75 cm telescope."* There are many instances in astronomy of telescopes sensitive to very different parts of the electromagnetic spectrum working together to achieve a common goal. This is not an exception, it is the norm, and should be carefully planned for. *"It regularly monitors VHE sources also visible in optical wavelengths, measures fluxes of H.E.S.S. sources in different colours simultaneously with the main array, triggers IACT observations in case of flaring activity, and provides atmospheric calibration data."* Such multi-messenger facilities are used for multiple purposes. *"H.E.S.S. is operated by shifts which rotate…made up of the scientific body of the…collaboration… This personnel usually has no experience in the operation of an optical telescope. … Since the beginning of operations in…2006, more and more subsystems facilitating robotic operations have been added… The robotic telescope was still limited in two ways: It required staff performing repeating tasks every day, and it could not react to any sort of event during the night."* If a multi-telescope, multi-bandpass facility is too-simply summed up as being an example of one (*e.g.,* VHE) instead of both (here, VHE + optical), then decisions of staffing, maintenance, software, data archiving, *etc*., will be biased by those expectations. *"[T]he need to avoid false positives usually requires a second observation… Without an observer present at night, such a follow-up can only take place during the next night, with H.E.S.S. finally observing the night after."* Coordinated observations between multiple facilities, especially in widely separated regimes of bandpass or methodology, can introduce interruptions into the observational workflow.

From Colomé, *et. al.* [27] [9149-17] *"The scheduling procedure ensures that long-term planning decisions are correctly transferred to the short-term prioritization process for a suitable selection of the next task to execute on the array. In this contribution we present the constraints to CTA task scheduling that helped classifying it as a Flexible Job-Shop Problem case and finding its optimal solution based on Artificial Intelligence techniques. We describe the scheduler prototype that uses a Guarded Discrete Stochastic Neural Network (GDSN), for an easy representation of the possible long- and short-term planning solutions, and Constraint Propagation techniques."* There is no scenario for the future of astronomy that does not include an ever greater role for software, and our publications will rely on increasing amounts of concepts and terminology from software engineering and related disciplines. It would be well to keeping in mind, however, what the eminent computer scientist Edsger W. Dijkstra had to say: "Computer science is no more about computers than astronomy is about telescopes." Two papers presented at this conference made the exciting claim of identifying an optimal solution to the notoriously difficult problem of telescope scheduling. These solutions are, however, different from each other – to the extent that completely different terminology (and apparently, formalisms) are used to describe them. *"A simulation platform, an analysis tool and different test case scenarios for CTA were developed to test the performance of the scheduler and are also described."* This being the case, a coherent test plan or data challenge and the tools needed to implement it, along with the necessary simulated or real test data sets, and evaluation criteria, *etc.,* are particularly important in a multiple observatory context.

Edwards, *et. al.* [28] [9149-18] says *"The Australia Telescope National Facility operates three radio telescope: the Parkes 64m Telescope, the Australia Telescope Compact Array (ATCA), and the Mopra 22m Telescope. Scientific operation of all these is conducted by members of the investigating teams rather than by professional operators. All three can now be accessed and controlled from any location served by the Internet, the telescopes themselves being unattended for part or all of the time."* An autonomous paradigm can extend the life of mature facilities and preserve community investment. Often this implies an operational transition from one staffing model to another. Needless to say the internet is often at the center of these capabilities, with all the strengths and weaknesses that implies. *"Observatory software to control and monitor observations is run within Virtual Network Computing (VNC) sessions. This allows seamless handovers between observers, and ensures that when local staff are contacted to provide advice, they are able to see exactly the same displays that the observer is viewing, and, if necessary, take control of the observations to address any problems. Experience has found that the use of VNC sessions is generally good, though not completely robust, as certain keystrokes or actions on an observer's machine cause VNC windows to lock-up or freeze on occasions, often requiring staff intervention."* VNC is one frequent technology choice for the reasons well stated here. And it has the fragility also mentioned. A coherent system engineering process includes a trade study of different off-the-shelf (OTS) technologies and will proactively identify areas of weakness that should be addressed. In particular, an observing environment that is well defined and well behaved locally to the telescope control room will always exhibit unintended behavior when run remotely, no matter how exact a replica the windows on the observing console appear to be. One area of particular concern is cyber-security. A large telescope facility halfway around the planet is an attractive target for malefactors with a certain adolescent mind set. The ease of use that VNC and other general-purpose remote technologies provide to authorized users is a possible entrée for unauthorized hackers. Major telescope facilities are not very different from industrial work sites; slewing a telescope may be a nearly silent operation involving many tons of equipment that can present a significant risk of injury. These are all issues to be addressed through systems engineering best practices.

Giordano, *et. al.* [29] [9149-15] state *"With the new generation of telescope…the management of this instrument does not take into account the optical quality of the atmosphere. Therefore, the observational time is not optimized and there are losses of time and money to the detriment of scientific advances. It is then important to improve the management by taking into account the optical conditions above a site. … it is interesting to look at the evolution of the seeing during the night. … one can conclude that better is the resolution of our domain better is the forecasting … Also the terrain model has an incidence on the simulation, and it should be a fine as possible to have a good estimation of the dynamic flow above the ground. Its influence should be more visible if the resolution of our domain was closer to the resolution of the orographic data."* Observational astronomy provides excellent use cases for a wide-range of applied science and engineering infrastructure. Nobody cares about the fine details of weather forecasting more than astronomers, and our community can drive significant improvements in the practice of meteorology. This is true in the fine details of weather local to a particular mountaintop as in this paper. And would also be true across a larger area of continental or island topography spanning several observatory sites. *"However, there are two main problems for these improvements: …The computational time would not be enough to use our model for the flexible scheduling. The second problem comes from the refinement of the terrain model. Indeed a good resolution lead to have a rougher terrain, and some instability for the simulation of the wind velocity could appear, stopping the simulation."* Benefiting from such efforts requires a commitment of large computational resources, but also faces inherent model-dependent computational issues. At some point an investment in what is to astronomers applied physics turns into its own research activity.

Shen, *et. al.*[30] [9149-90] says *"[The] ALMA software…is ready for remote operation and very few adjustments have to be introduced. The most critical technical aspect is the centralized file system, which is shared across multiple servers via NFS. This should be avoided in the remote site… subsystems must be configured to use the local copy of the database… such flexibility in configuration should be implemented in the future releases"* Most certainly an operations site should avoid mounting NFS partitions remotely. However, by introducing the concept of working from a local copy of the database, a requirement to synchronize the two copies is revealed. This implies the existence of privileged users, of a procedure or script implementing the synchronization process, of a schedule to adhere to, and of documentation and recipes for cleaning up the inevitable problems that will result. *"Having two control rooms to be able to control the same array could be problematic if both sites were not well coordinated. Therefore we recommend having a permanent audio and video communication in both sides, and a new protocol of operation should be established"* Having two control rooms is similar to having two VNC windows. One feature of VNC is the ability to run it in read-only mode. Perhaps something similar could be implemented for the control room as a whole. Alternately one is left trusting personnel at one site to avoid typing (or mis-typing) a contradictory command. An open duplex A/V link is a good idea, but is known to generate negative feelings from some staff. *"[W]e observed a very positive effect in the staff which works in Santiago and has had very few opportunities to get involved in the day-to-day operation of the array. The remote operation will also create an opportunity to the organization in term of saving in the operational costs, and more importantly, it will also reduce the overall ecological footprint of the project."* There are indeed numerous positive benefits possible from introducing a new operations paradigm.

From Rahmer, *et. al.*[31] [9149-86] *"The Federal Aviation Administration (FAA) has the authority to regulate the safe and efficient use of the navigable airspace, and to that end it requires that laser operators follow a process of registration and notification of planned laser operations. … [T]he observatory is required to coordinate laser operations with the United States Air Force's "Laser Clearinghouse" (LCH), which reviews all proposed laser illuminations, provides predictive avoidance and safe laser operating parameters, coordinates with satellite owner/operators (SO/O) and mission partners, and when notified, reports all laser activities inadvertently conducted outside authorized parameters. … Final activation of the propagation system is done by turning a key … A replica of the laser warning lights exists in the telescope control room… [T]he laser operator station in the control room has a laser e-stop button. … Several external agencies need to be informed prior to the beginning of laser operations[:] VATT (Vatican Advanced Technology Telescope), which is one of the three telescopes on site … MGIO (Mount Graham International Observatory). The director of MGIO is also informed in advance, and the notification includes the University of Arizona Police Department (UAPD) … FAA. The airspace above Mount Graham falls under the control of the Albuquerque Traffic Control Center. Following our notification, the FAA releases a NOTAM (NOTice to AirMen) … JSpOC (Joint Space Operations Center), which is the office responsible of the day-to-day coordination for satellite avoidance (in conjunction with the LCH). They require daily contacts at the beginning and at the end of operations. … JACKAL Military Airspace Manager. Mount Graham is inside an airspace area used for military aircraft training from the Tucson-based United States Air Force Base."* The overlapping institutional areas of responsibility here are remarkable, and it may simply be that some astronomical observing modes are too complicated to be automated. However, it is precisely the large number of stakeholders that argues for a coherent planning process and perhaps a community-wide response on issues like this. Another Laser Guide Star example follows; it is interesting to compare the similarities and differences.

Marin, Cardwell and Pessev[32] [9149-79] say *"On a normal night at Gemini two individuals operate the telescope and the instruments… Due to the added complexities of running GeMS [Gemini Multiconjugate adaptive optics System] there are a significantly higher number of people on the summit. Currently the minimum number of people needed to run GeMS is seven… In addition to the nighttime personnel on the summit there are additional personnel on call… Gemini sends a request to the LCH 4 business days before a planned night of LGS operations. … Twenty-four hours prior to the start of our LGS night LCH sends us a list of approved targets with timing windows when we are allowed to propagate our laser. This text file is not very user friendly… A team of three spotters is used on a GeMS night, working in one-hour shifts. One spotter is outside physically looking for airplanes, one is inside monitoring a radar system (VITRO) and the third is on recovery time. The advantage that Gemini south has over normal aircraft spotting done by individuals outside is VITRO. This is a radar system provided to us by the Direccion General de Aeronautica Civil (DGAC) the Chilena equivalent of the US FAA. … A GeMS acquisition is the most complicated currently done at Gemini South. It requires that several different people do various tasks in parallel… While the internal complexity of GPI rivals*

*that of GeMS, the GPI instrument was designed to be run by a normal Gemini night crew… It is expected that by the end of 2015 the nighttime staff [for GeMS] will be reduced to five people. … The AO specialist will be removed from the summit and the task of loop optimization will be moved over to a combination of the telescope operator and the queue observer. The AO specialist will now be callable support… There will also be a reduction in the number of aircraft spotters from three to two. This is accomplished by removing the need for an external aircraft spotter, as well will rely solely on radar information and aircraft transponder signals. From 2016 to 2018 the number of nighttime staff will be further reduced to three. Again this will be accomplished by fully removing the aircraft spotters, all aircraft detection will be done via software. By 2018 we reach our final operation mode in which there are only two people at the summit, which is the same as the current non-GeMS operations. The laser operator role will be transferred over to the telescope operator. Of course this will require a large software and hardware effort to ensure that we do not add increased overheads…"* And on the other hand, some combination of the appropriate technogies (*e.g.,* here the VITRO radar system) and carefully defined procedures may permit more automation for some specific complex systems than might be predicted.

Carroll, *et. al.*[33] [9149-12] *"We propose a drastic change to the scheduled operation of LSST in order to increase cadence and to improve Type Ia SNe capture for dark energy studies. … The astronomical community will benefit from the information LSST collects over its ten-year survey. It is imperative that this survey be optimized to produce maximum results. With a multitude of science drivers, a balance must be struck between competing demands for cadence and survey uniformity. … This study illustrates the need for more in-depth analyses of the effects large dithers have on each of the science objectives. … We caution the reader that our estimates of the yield of Type Ia supernovae are highly approximate and result from an analysis of the total number of LSST observations without regard to which filter those observations were made in."* 'Drastic' is a drastic choice of words, but it was the word used by the authors themselves. In general, different stakeholders will express significantly different perceptions of the science and engineering requirements for community projects. This is normal and healthy. The meta-requirement is for a clear and open requirements discovery process that stakeholders agree to in advance. Such a process will indeed be greatly aided by pertinent research analyses and by clearly establishing the limitations of various estimates.

From Vitale, *et. al.*[34] [9149-6] *"Beside the obvious interest that a direct GW detection deserves on its own, there is much interest in the possibility of joint electromagnetic – gravitational detections."* Coordinated observing modes address some of the most scientifically intriguing questions in astronomy. *"The LIGO-Virgo Collaboration is working toward providing data quality information in low-latency in the advanced detector era and fully automating the validation process. Triggers surviving the final validation are assigned a unique ID and added, together with some relevant information…to the gravitational-wave candidate event database (GraceDB). … Skymaps of short GW bursts emitted by other kinds of astrophysical sources are generated using cWB (coherentWave Burst, a Poisson point estimate), with typical latencies of <~ 1 minute."* **Addressing the science use cases requires speed. Speed requires automation of complex workflows.** *"Once the sky map is available on GraceDB, a robot checks if the FAR [False Alarm Rate] is below the threshold LIGO and Virgo fixed to proceed with EM follow-up."* Throughout astronomy, but especially in the time domain, it is not good enough to infer the best answer – in addition an estimation of the likelihood of error is also needed. This requirement increases dramatically when external facilities may be following up on the original observations. *"For triggers which pass this criterion, an alert will be sent to the partner EM facilities. A VOEvent-formatted notice will be circulated through a private version of GCN, also containing a link to a FITS file sky map. Each EM facility may decide to further downselect the events to follow, if the threshold FAR fixed by the LIGO-Virgo Collaborations resulted in impractically frequent observations."* A standardized VOEvent [REF] message is published to a pre-vetted list of subscribing projects. This message references a database of auxiliary information. Individual subscribers retain the right to respond as they desire. If alerts are 'impractically frequent', all subscribers may respond only to the highest likelihood events, duplicating efforts and leaving many alerts without follow-up. **Rather, coordination of follow-up observations is required.**

## 2.1 Questions to consider

This leads us to the remainder of the paper. How does an "EM facility" (that is, an astronomical observatory) make decisions about pursuing scientific opportunities, especially related to celestial transient events? What strategy maximizes the likelihood of a positive outcome? How are multiple opportunities evaluated one to the next? What is required for efficient response to random occurrences? How should two EM observatories best compete – or cooperate – to maximize the joint science accomplished between them? What community infrastructure is needed to permit community-wide participation in such science? What data and software standards need be adhered to? How does astronomy avoid a *Tragedy of the Commons*[35] "overgrazing" scenario, in which the utility to individual investigators or observatories of focusing each only on the highest-priority alerts rationally compels the community to ignore the great majority of all other alerts?

## 3. A DYNAMIC COALITION BROKER

Time domain astronomy is now moving into an era when the quantitative number of potential target candidates will force a qualitative change in our observational approach to the study of transients. Particularly stressed will be the study of fast transients with durations of hours or less. The current state-for-the-art for the follow-up of fast astrophysical transients is exemplified by the robotic telescopes that rapidly respond in real-time to gamma-ray burst triggers (*e.g.,* Akerlof, *et. al.*[36] 2003, Vestrand, *et. al.*[37] 2002, Klotz, *et. al.*[38] 2008). Those telescopes can respond and begin observing in as fast as 5 seconds after receipt of a transient localization message. However, the multi-wavelength follow-up of GRB triggers still employs a "second grade soccer" approach – all the follow-up telescopes respond in a static pre-scripted manner ("pounce and kick at the ball") that doesn't assign roles or coordinate the overall response. Instead, the GRB follow-up telescopes respond robotically and for the most part co-ordination of the larger follow-up response is conducted by humans through messages posted to the GCN. This approach has worked well for the low rate event rate of a few per night for GRBs, but it will not scale to event rates of a hundred per night. And at rates of a thousand or more transients per night, if the transients are not ranked in importance and follow-up instruments are not co-ordinated, the scientific return employing current GRB follow-up approach will collapse in a manner reminiscent of an internet denial of service attack.

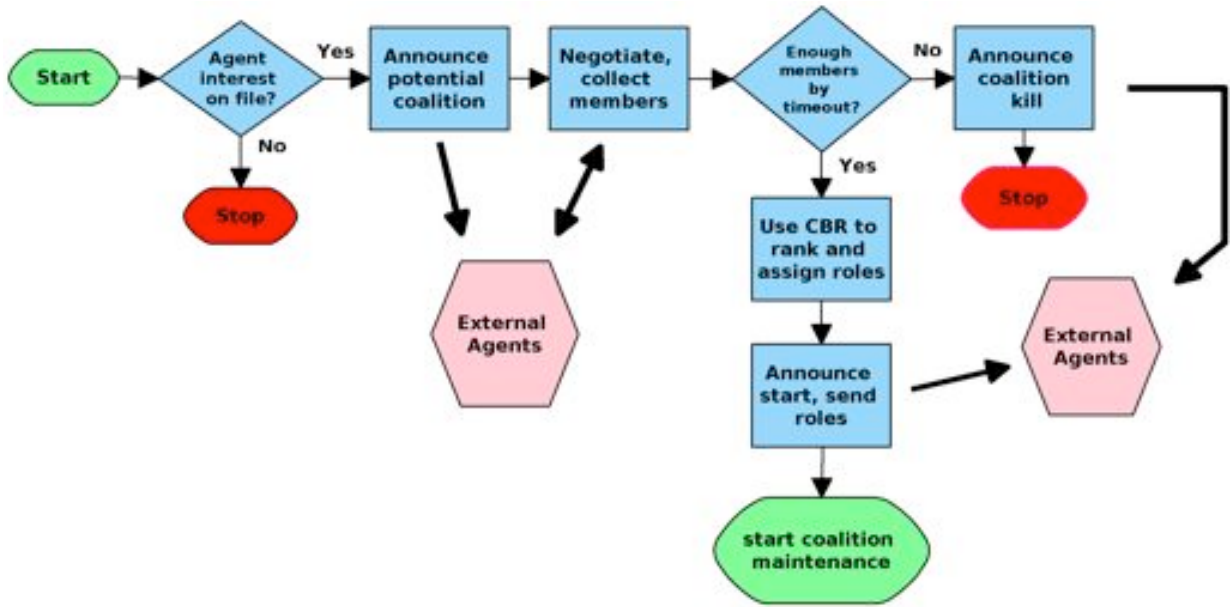

**Figure 1.** The initiation of a dynamic coalition is executed after detecting a new transient event. A broker is managing the offered participations, ranking them based on capabilities, and assigning roles automatically.

A promising approach to the coordination of multi-wavelength follow-up problem at high event rates is to employ techniques that are being developed for distributed artificial intelligence. The two poles of distributed artificial intelligence approaches are: (1) distributed problem solving and (2) multi-agent systems. The simplest distributed problem solving approach divides the problem among distributed identical "nodes" which are centrally designed and controlled as slaves. The other extreme is the multi-agent approach that employs autonomous intelligent agents with heterogeneous capabilities and no central control. Agents are self-motivated and act only according to their own success criteria and can generate inefficiencies in overall system performance. This approach resembles the current approach employed by Astronomers. Optimization of the overall scientific return from robotic instruments in an opportunity-rich environment would benefit from a hybrid approach that blends the multi-agent approach with elements of central control that would assign roles that are dynamic as collective knowledge of the nature of the transient evolves during the follow-up. A Dynamic Coalition Architecture is an AI approach that would provide those benefits.

The central idea in the Dynamic Coalition Architecture is to employ temporary partnerships between scientific assets to optimize the scientific return on the study of a given transient. Here each new transient is spawned as a potential coalition and is distributed by a transient broker in real time to a collection of external agents with an invitation to join the coalition. Some of the key information included in the participation request is the classification of the transient and a measure of certainty of the classification, the celestial coordinates, the age of transient, the source of the transient trigger, *etc*. The broker collects coalition members and determines before the expiration of an opportunity window (which is event class dependent) if sufficient interest is present to merit an organized coalition follow-up. If not, the potential coalition is killed and agents that expressed interest are notified. But if sufficient interest is present, the start of the coalition is announced and participating agents are assigned roles that reflect their capabilities. Figure 1 illustrates the coalition initiation process.

The assignment of roles by the broker coalition manager is based on meta-data about the external agent. Important factors include geographic location, agent capabilities and potential configurations, as well as current state and real-time reconfiguration capability. Another important factor is agent effectiveness as judged by previous coalition behavior and the quality of the results. The management of a coalition is illustrated as a flow chart in figure 2. It is important to recognize that coalition agents are not limited to representation of just telescopes or instruments. Agents could also represent robots (or humans) mining an archive of historical information or a theorist performing a needed calculation.

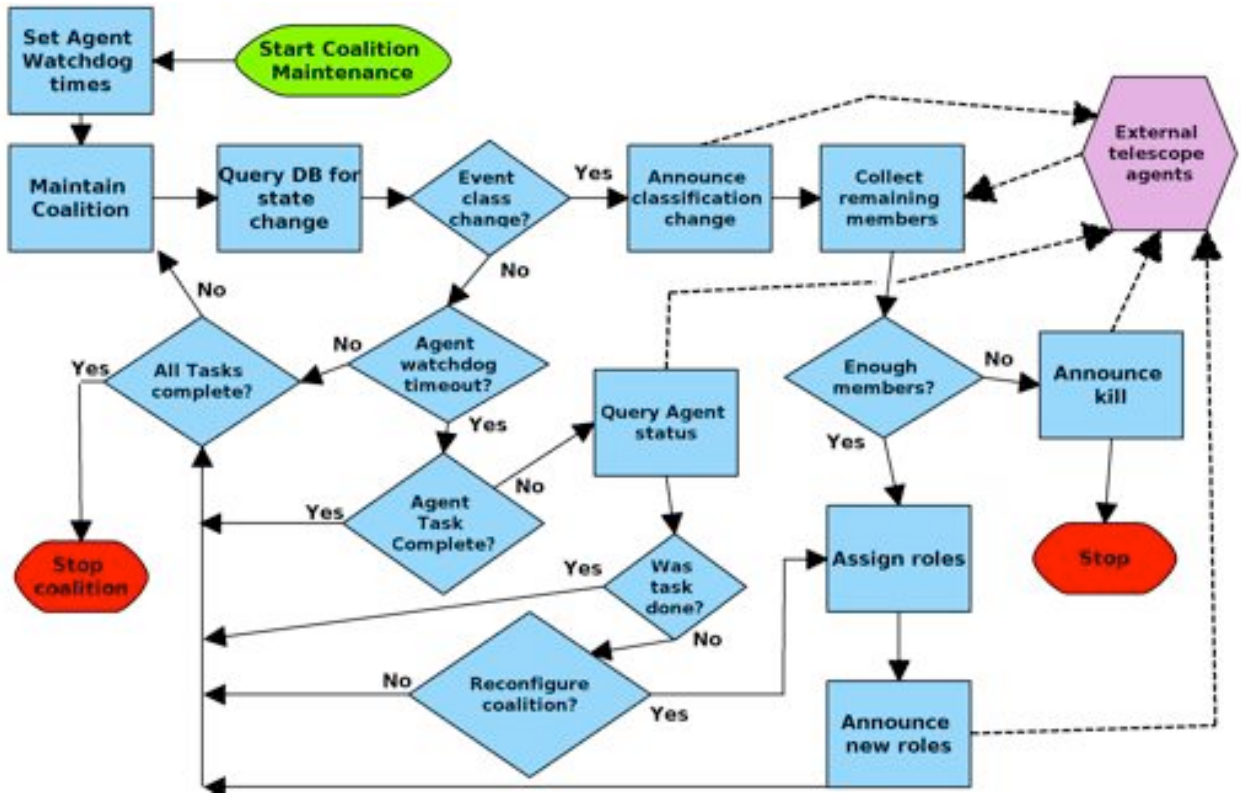

**Figure 2.** A flow chart outlining the management of a Dynamic Coalition.

An important consideration for the coalition manager when assigning roles is the classification of the transient and the nature of the execution plan for optimal follow-up of a transient of that type. A practical approach for follow-up in during early generations of the system would be to employ Case-Based Reasoning with Learning. Here a case includes a transient description and its current state – *e.g.*, a fresh Swift-localized Gamma-ray Burst, a LIGO gravitational wave event, a fading SN Ia, unknown bright optical transient, *etc*. And it includes a solution description that describes what measurements need to be made and when they need to be made to collect the key scientific information.  After conducting the case-based follow-up it will be important to assess the quality of the solution outcome. This feedback will allow human subject matter experts to improve the solution and allow the system to learn.

We are entering an exciting new era of time domain astronomy where there will be an overwhelming number of transients found in real-time. Coordinated multi-wavelength follow-up will be essential for optimizing the scientific return from transients that will be diverse and range from multi-messenger events like Gravitational Wave events to electromagnetic transients over the full spectrum from gamma-rays down to radio frequencies.   A multi-agent approach employing a Dynamic Coalition Architecture has many attractive aspects that would be useful for optimizing follow-up by the world's heterogeneous collection of instruments.  To some extent, this approach will require a reengineering of observatory operations and the adoption of internationally standardized agent negotiation protocols and communication language.

## 4.  A TOOLKIT FOR AUTOMATED FOLLOW-UP

Having a responsive network like that described in the previous section is a scientific dream that can only be realized by addressing very practical questions and defining concrete ways of implementing a workable system and identifying paths whereby such a system might be adopted by a significant number of participants.  The most logical path is the creation of a "toolkit" which can be adapted to local software, reducing the administrative and software overhead required at each site and reducing the sociological problem to a question of the amount of local resources required to adapt the local system to a given global infrastructure.   Thus, four very practical software / sociological / institutional problems must be solved:
1. Deciding what concrete scientific/technical information needs to be passed (the actual protocol – that is the detailed schemata expressed via a format like XML – is just a technical detail).  This would appear to be obvious for a given scientific goal: *e.g*., if one wants to permit global imaging, then the trivial constraints of aperture, bandpass, location, and availability are easily expressed, for example by using RTML.[39]  However, as soon as one is concerned with the quality of the data, these issues become more complicated. The use of documented automated calibration pipelines should alleviate much of the latter problem.  Given modest initial goals, this should not be a serious problem.
2. Deciding what transaction model is needed: this is non-trivial because the resources of a global heterogeneous network are – by definition – constantly changing, for example an intended service requested earlier by a client and promised by a telescope server may in fact never be performed due to weather.  The more transaction overhead visible to the telescope servers, the larger the chance that collisions will occur with local constraints and the more software that needs to be written locally.  On the other hand, a complex transaction agent hidden from the end-user by the toolkit may be exactly the added benefit needed to convince a resource to join the network.  One model, that of a "dynamic coalition broker", was described in detail in the previous section and much of its complexity could be hidden behind the façade of the toolkit.
3. Reducing the complexity of the content and the number of explicitly present agents and transactions visible to the local systems to a minimum, thereby maximizing the chance that a potential participant would spend local resources to obtain compliance.  The solution must be generic enough that it could be adopted by a large number of willing players for different scientific reasons but not so generic that the players are forced to expend lots of resources to implement the local parts needed but not explicitly provided by the toolkit.
4. The system must be easily configurable for different, hopefully parallel running, projects.

Since what is required is a layer between a purely global infrastructure and a purely local infrastructure, each local solution will have to be tailored individually to some extent.  Thus, the maximum benefit for the global system would be to target local systems using identical or similar infrastructures.  There is a very wide variety of automated telescope and observatory control software systems operating scientifically useful telescopes with an equally large range of apertures, capabilities, and complexities.  For historical, technical, sociological, institutional, as well as financial reasons, it is unlikely that currently operating automated telescopes would be re-fitted to use a system more conducive to global networked operations (although not unthinkable, for example the recent software-refitting of MONET[g] to the STELLA system). Thus, a communal toolkit for enabling global networking can most pragmatically offer tools for the handing of the external protocol (*e.g.*, parsers), best-use examples, and strategic implementations that work practically "out-of-the-box".

---
[g] https://monet.uni-goettingen.de

Fortunately, current observatory systems can easily be separated into those with few installations and/or accessibility issues, which are then unlikely to be integrated at first, and those with many installations, active support for the integration, and/or with institutional/scientific pressures. There are currently two major systems which have been implemented on a large number of heterogeneous telescopes, mainly ASCOM[h] (60+ telescopes running Windows software) and RTS2[i] (>20 telescopes running Linux). Large and administratively homogeneous observatories like ESO may use a homogenous internal system such as BOSS[40], but the non-homogenous parts may be too complex and numerous and administrative resistance too great, despite the potential impact of opening large, well-instrumented telescopes to a global network. There are also diverse systems running small numbers of telescopes, such as Audela[j] (about 10 installations), STELLA (~5 instances), and INDI[k] (~10 instances?). These are likely to adopt the toolkit simply because they are interested in the same concept and since doing so is aligned with their missions.

## 5. TRANSIENT ALERT INFRASTRUCTURE

The astronomical time domain is in a period of unprecedented growth. Even a narrow inventory of facilities such as "ground-based optical sky surveys" reveals a long list with current and near-term highlights like Pan-STARRS[l], Catalina Sky Survey[m], Zwicky Transient Factory[n], and ramping up to LSST[o]. Few such surveys attempt their own comprehensive follow-up on the celestial transient events they discover. The asymmetry in aperture-size requirements and targeting efficiency between a wide-field photometric survey and the complementary spectroscopic follow-up observations creates the well-known deficit in the latter class of facilities.

### 5.1 Distributing the load

For some science use cases a survey can provide its own follow-up, limited to the accumulation of time-series data in the filters installed in one particular camera, and with phases constrained to the diurnal observing limitations of a single mountaintop. On the other hand, for certain high-value classes of events, most obviously Gamma-Ray Bursts (GRBs), multi-telescope spacecraft / facilities work in close coordination to localize and follow-up on targets. In all other instances notifications of celestial transient events, using the widely-supported VOEvent[p][41] protocol for example, must be circulated either publicly or privately to other telescope facilities.

A Dynamic Coalition Broker (DCB) as discussed in section §3 is required to automate negotiations over which facility should follow-up which transient alerts. The DCB addresses both the issue of efficient rapid-response follow-up, but also the question of efficient use of observatory facilities. In the absence of some central arbiter such as the DCB, all issues of the allocation of observatory resources must be addressed by human-mediated policy discussions, either very coarsely conceived on a per-observing-semester timescale (or longer), or very finely argued over the relative merit of a single transient alert. Similarly, the automated follow-up toolkit described in section §4 will provide tools that can be used to maximize both the efficiency of alert follow-up as well as the efficient utilization of observatory resources. Since the observations made with any mix of instrumentation on any telescope include both images of the static sky as well as of varying phenomena, these gains then apply to the resulting productivity of the participating facility as a whole. Rather than competing with other programs for telescope time, a well-managed coalition optimizes the resulting observations, whether carried out via queue scheduling or some modified classical observing paradigm.[42]

---

[h] http://www.ascom-standards.org
[i] http://www.rts2.org
[j] http://audela.org
[k] http://indilib.org/devices/telescopes.html
[l] http://pan-starrs.ifa.hawaii.edu
[m] http://www.lpl.arizona.edu/css/
[n] http://www.ptf.caltech.edu/ztf/
[o] http://www.lsst.org/lsst/
[p] http://voevent.org/

Neither the automated-astronomy toolkit or the Dynamic Coalition Broker mandate any particular observatory policies over access to individual facility. It always remains in the hands of particular observatories to decide whether to participate. However, the more aligned a facility becomes with community standards and the overall engineering of the system, the greater the possible gains both locally and globally. The user community for each facility grows, just as the diversity of facilities open to individual members of the astronomical community ramifies in a network effect.[q] Such autonomous technologies reaching a critical mass is a prerequisite for future market-driven strategies to take hold.[43]

## 5.2 Closing the event loop

The ultimate goal of re-engineering astronomical infrastructure is to close the event loop such that a discovery alert (time-domain or otherwise) autonomously triggers follow-up observations, archival searches, pipeline processing, theoretical simulations, *etc*., that themselves generate new alerts and the scheduling of new observations. The alternative would be to imagine that one or more of the individual steps above is limited to manual human methods. Note that the actual observations might well be carried out by human astronomers; a particular telescope or a finicky instrument might benefit from expert attention – but what happens to generate an observation request on input or the data handling on output most certainly can be automated even in those cases. Such a system does not remove human judgment from the loop, but rather provides the power tools for humans to most efficiently supply their judgment and to do so at the most efficient points in the process.

Systems engineering requirements for infrastructure enabling a collection of facilities to interoperate include:

- ubiquitous network connectivity to remote sites
- widespread adherence to data standards (VOEvent[r], FITS, ...)
- but with flexibility for special cases
- sensitivity to esoteric issues like data compression[44,45] and timekeeping[46]
- easy (re)configurability
- autonomous operation, but
- friendly to humans and benefiting from those tasks humans do best (for instance, many pipelines include human review as an explicit step in the workflow)
- efficient command and control (see §3)
- autonomous technologies and protocols (see §4)
- attention to community-building[s] [47]
- specific science goals[48] – it is ultimately all about experimental design

Figure 3 shows how the Dynamic Coalition Broker (DCB) and ANTARES broker can be used to close the event loop among facilities networked with an automated-follow-up toolkit and VOEventNet. This is overlaid on a diagram of LSST transient alert workflow[49] presented in this conference at SPIE 2010. The blue circles highlight various subsystems of the community. In addition to systems operated by various observatory-level entities is infrastructure common to all, in particular the network for distributing VOEvent alerts, and the parallel network for reaching optimal decisions regarding which facilities will follow-up on which alerts. ANTARES is an example of a value-added service improving one or more event streams. The flow proceeds from publishing of the raw LSST events (either at the base center in Chile or the archive center at NCSA, #s 1, 2 & 3); followed by 4) ANTARES (or other brokers) filtering the highest priority events; then these events pass through to 6) VOEventNet, and to the DCB for negotiation between follow-up facilities as described in §3; efficient 5) follow-up, and 7) value-added services such as enrichment of the event streams from archival data products can then occur. Individual observatories remain in control of their own assets, for instance the robotic telescope at the bottom can accept either its own scheduling decisions, or these can flow from VOEventNet and DCB. The diagram is complex and rather crowded, but this is true whether or not the organization into a system-of-systems is recognized, whether or not the alerts of celestial transient events are handling autonomously via VOEventNet, and whether or not the scheduling of follow-up observations is done autonomously using some variation of a dynamic coalition infrastructure. Clarifying the systems and deploying infrastructure makes for efficient operations.

---

[q] http://en.wikipedia.org/wiki/Network_effect
[r] http://www.ivoa.net/documents/VOEvent/
[s] http://hotwireduniverse.org

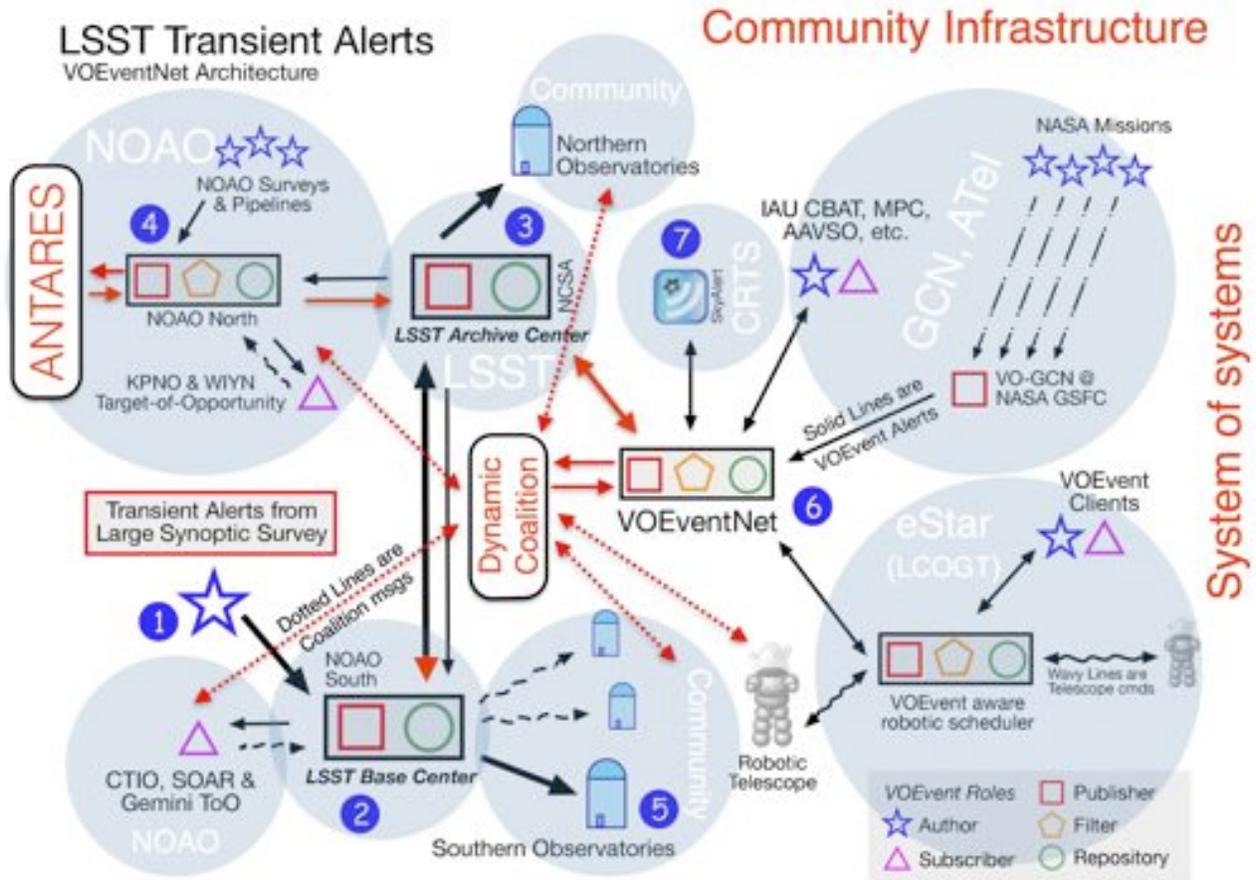

**Figure 3.** ANTARES and Dynamic Coalition brokers overlaid on diagram of LSST VOEvent flow presented at SPIE 2010. Red arrows show flow of characterized events from ANTARES triggering coordinated coalition building and the resulting follow-up observations. These may dynamically feed new VOEvents back into the network, closing the loop.

**5.3 Real-world constraints**

In 1888, eminent astronomer Simon Newcomb is said to have said, "*We are probably nearing the limit of all we can know about astronomy.*" More than a century later we recognize that we are nowhere near fulfilling his prophecy, but perhaps we are beginning to ask the right questions. How to answer those questions is an exercise in experimental design, data- collection, management & mining, and the contingent decision-making that ties it all together.

In 1976, eminent computer scientist Leslie Lamport's paper discussing the computer arbiter problem, "*On the Glitch Phenomenon*"[t] was rejected by *IEEE Transactions on Computers*, who presumably took exception to his making the case that computer science could never eliminate all the glitches and ghosts from its machinery. He then famously (well, famously in that community) wrote a paper, "*Buridan's Principle*"[u], pointing out that the arbiter problem occurs in everyday life. It took him an additional twenty-eight years to get this second paper published[50], being turned down along the way by *Science* and *Nature*. What is the arbiter problem?

---

[t] http://research.microsoft.com/en-us/um/people/lamport/pubs/pubs.html#glitch
[u] http://research.microsoft.com/en-us/um/people/lamport/pubs/pubs.html#buridan

> "*Buridan's Principle.* A discrete decision based upon an input having a continuous range of values cannot be made within a bounded length of time."

An inherent race condition of the universe, Buridan's Principle (named for a proverbial ass that starves halfway between two bales of hay) is encountered whenever a traffic light turns yellow as a driver approaches the intersection. Should I stop or should I go? And it provides a sufficient explanation for how cars end up being struck by trains even though both the decision to stop and wait for the train to pass before crossing the tracks, and the opposite decision to gun the engine and race the train to the crossing would have been safe. Astronomical systems are no more subject to glitches than others – but are also no less so – while our parameter space is literally universal in scope, and our systems and technology are frequently unprecedented in their requirements.

Given the lengthy lifespan of astronomical facilities, the topic often arises of a need for reengineering some telescope, instrument, software or system to meet a scientific opportunity unpredicted by Newcomb. Our community's most comprehensive efforts[v] yet fall short of pursuing an engineering agenda for the entire astronomical system of systems[51], though this would be the most efficient use of resources to reach our ambitious scientific goals. The time domain provides both the motivation and means to begin incrementally building a pervasive autonomous scientific architecture.

---

[v] http://www.ivoa.net